\documentclass[english,prl,aps,reprint,longbibliography]{revtex4-2}
\usepackage[T1]{fontenc}
\usepackage[latin9]{inputenc}
\usepackage{amsmath}
\usepackage{amssymb}
\usepackage{graphicx}

\makeatletter

\newcommand*\LyXThinSpace{\,\hspace{0pt}}

\makeatother

\usepackage{babel}
\begin{document}
\title{Signature of parity anomaly in the measurement of optical Hall conductivity
in quantum anomalous Hall systems}
\author{Zi-Ang Hu, Huan-Wen Wang, Bo Fu, Jin-Yu Zou, and Shun-Qing Shen}
\email{sshen@hku.hk}

\affiliation{Department of Physics, The University of Hong Kong, Pokfulam Road,
Hong Kong, China}
\begin{abstract}
Parity anomaly is a quantum mechanical effect that the parity symmetry
in a two-dimensional classical action is failed to be restored in
any regularization of the full quantum theory and is characterized
by a half-quantized Hall conductivity. Here we propose a scheme to
explore the experimental signature from parity anomaly in the measurement
of optical Hall conductivity, in which the optical Hall conductivity
is nearly half-quantized for a proper range of frequency. The behaviors
of optical Hall conductivity are studied for several models, which
reveal the appearance of half-quantized Hall conductivity in low or
high frequency regimes. The optical Hall conductivity can be extracted
from the measurement of Kerr and Faraday rotations and the absorption
rate of the circularly polarized light. This proposal provides a practical
method to explore the signature of parity anomaly in topological quantum
materials.
\end{abstract}
\maketitle

\section{Introduction}

Parity anomaly describes the fact that a single massless Dirac fermion
in 2+1 dimensions undergoes a spontaneous symmetry breaking when it
is coupled to the $U(1)$ gauge field \citep{Redlich1984prl,Niemi983prl,Gaume1985physics}.
This anomalous effect is physically manifested as the half-quantized
Hall conductivity in the external electromagnetic field. Several condensed
matter systems have been proposed to simulate parity anomaly on a
lattice, such as a monolayer graphite \citep{Semenoff1984PRL,Haldane1988Model}
and PbTe-type narrow-gap semiconductor with an anti-phase boundary
\citep{Fradkin1986PRL}. In addition, the massive Dirac fermions break
the time-reversal symmetry and parity symmetry explicitly. At the
half-filling, the finite Dirac mass also leads to a half-quantized
Hall conductivity \citep{Haldane1988Model,Lapa2019prb,Nielsen-81plb,Burkov2019prb}.
Combined with the parity anomaly, the massive Dirac fermion exhibits
an integer-quantized Hall conductivity, which leads to the quantum
anomalous Hall effect in condensed matter systems \citep{yu2010science,Chang2013sci,Chu2011prb,wang2015Quantized,Liu2016RevCM}.

As the contribution from the parity anomaly and Dirac mass are mixed
together, it is difficult to distinguish the two mechanisms from the
total Hall conductivity in the dc limit. In 1988, Haldane proposed
that the half-quantized Hall conductivity from the parity anomaly
could be realized if an unpaired Dirac fermion appears at a critical
transition point between a normal insulator and a Chern insulator
phase \citep{Haldane1988Model}. In recent, another attempt was reported
in the semi-magnetic topological insulator thin film, where only the
top surface state was gapped by the magnetic doping, and a nearly
half-quantized Hall conductivity was observed \citep{Mogi2022NP}.
Nevertheless, it is known that a single symmetry-protected Dirac fermion
does not exist on a two-dimensional lattice \citep{young2015prl}.
Hence, it is desired to explore a new method to figure out the half-quantized
Hall conductivity from the parity anomaly from a system where a single
gapless Dirac point cannot be realized. Fortunately, the physical
origins of two half-quantized Hall conductivities of the massive Dirac
cone are different, one is attributed by the low energy fermions around
the Dirac cone and the high energy regulator part respectively. Thus
they can be distinguished at different energy scales by optical Hall
conductivity.

Recently, Tse and MacDonald proposed that Hall conductivity at finite
frequencies can be detected using the magneto-optical technique. This
is mainly reflected in that the Kerr and Faraday angles can be experimentally
implemented to detect the optical Hall conductivity of the system\citep{Tse2010prl}.
However, this series of work has mainly focused on the low energy
region of the Dirac cone, and the contribution of quantum anomalies
from the high energy region has yet to be explored\citep{Tse2010prl,fu2021prr,Wu2016Sci,liu2020npj,OkadaNC2016,Morimoto2009prl,Tutschku2020prb,Tutschku2020prr,Dziom2017}.
Meanwhile, the magneto-optical effect is naturally suited for exploring
response patterns in the high-energy region, which provides us a possible
way to detect the signature of parity anomalies.

In this paper, we propose a method to detect the signature of parity
anomaly in a condensed matter system and to distinguish different
origins of the anomalous quantum Hall effect. We first calculate the
optical Hall conductivity of the Wilson fermions and massive Dirac
fermions analytically and get the expression of half-quantized Hall
conductivity by making Taylors expansion at the proper frequency.
Besides, we calculate the Hall conductivity of different lattice models,
including the Bernevig-Hughes-Zhang (BHZ) model, the Haldane model,
and the magnetically doped topological insulator thin films. Finally,
we discuss how this phenomenon can be implemented experimentally and
the effect that temperature and disorder can have on this, and we
propose that the character of this optical Hall conductivity can be
measured by several magneto-optical effects.

\section{Model Hamiltonian}

We begin with the two-dimensional Wilson fermion model \citep{Wilson-75,Rothe}

\begin{equation}
H=v\hbar(k_{x}\sigma_{x}+k_{y}\sigma_{y})+(mv^{2}-b\hbar^{2}k^{2})\sigma_{z},\label{eq:continuum model}
\end{equation}
where $v$ is the effective velocity, $k_{i}$ with $i=x,y$ are wave
vectors, $k^{2}=k_{x}^{2}+k_{y}^{2}$, and $\sigma_{i}$ with $i=x,y,z$
are the Pauli matrices. $2mv^{2}$ is the band gap at $k=0$, $b\hbar^{2}k^{2}$
is the dynamical mass regulator. In the dc limit, when the chemical
potential is located with the band gap, i.e., at half filling, the
Hall conductivity of the system is 
\begin{equation}
\sigma_{xy}=\frac{1}{2}\frac{e^{2}}{h}[{\rm sgn}(m)+{\rm sgn(b)}]
\end{equation}
\citep{shen2012topological,lu2010prb,Qi2011rmp}. Either the band
gap $mv^{2}$ and the regulator $b\hbar^{2}k^{2}$ contribute $\frac{1}{2}\frac{e^{2}}{h}$
to the Hall conductivity, which only depends on the signs, not value
of the two mass terms. When $m$ and $b$ have the same sign, i.e.,
$bm>0$, the Hall conductivity is quantized to be one in the unit
of $e^{2}/h$, and the system is is topologically non-trivial. When
$m$ and $b$ have the opposite signs, i.e., $bm<0$, the Hall conductivity
is equal to zero, and the system is is topologically non-trivial.
In the absence of the regulating term $b\hbar^{2}k^{2}$, the Hall
conductivity is $\sigma_{xy}=\frac{1}{2}\frac{e^{2}}{h}{\rm sgn}(m)$,
which contradicts the Thouless-Kohmoto-Nightingale-Nijs (TKNN) quantization
rule in a gapped system \citep{TKNN}. This means it cannot exist
on a two-dimensional lattice. The presence of the regulator $b\hbar^{2}k^{2}$
which provides another half-quantized Hall conductivity as $\frac{1}{2}\frac{e^{2}}{h}{\rm sgn}(b)$,
is essential to avoid the contradiction. When $m=0$, and the band
gap closes, the half-quantized Hall conductivity from the regulator
$b\hbar^{2}k^{2}$ can exist alone \citep{Fu2022qas,zou2022pas}.
In the case the parity is broken by the presence of he regulator $b\hbar^{2}k^{2}$.
However, when $b\rightarrow0$, the parity symmetry is restored, and
the Hall conductivity is still equal to $\frac{1}{2}\frac{e^{2}}{h}{\rm sgn}(b)$,
not zero as expected in the parity symmetry. This is the so-called
the parity anomaly in the lattice gauge theory \citep{Rothe}.

In the dc case, the Hall conductivity is equal to one or zero. We
cannot distinguish the contribution from the he band gap $mv^{2}$
and the regulator $b\hbar^{2}k^{2}$. As the two terms dominate the
low energy region ($k\rightarrow0$) and high energy regime ($k\rightarrow+\infty$)
separately, the two parts will respond disparately to an incident
electromagnetic field with a finite-frequency. Thus the optical Hall
conductivity may provide a possible way to distinguish the contribution
at different energetic scales.

\section{Optical Hall conductivity}

In this section, we will present the optical Hall conductivity at
a finite-frequency. In general, the optical Hall conductivity $\sigma_{xy}(\omega)$
at finite-frequency $\omega$ can be evaluated from the Kubo formula
\begin{align}
\sigma_{xy}(\omega) & =ie^{2}\hbar\int\frac{d^{2}k}{(2\pi)^{2}}\sum_{m,n}\frac{v_{mn}^{x}v_{nm}^{y}}{\epsilon_{m}-\epsilon_{n}-\hbar\omega+i\delta}\frac{f(\epsilon_{m})-f(\epsilon_{n})}{\epsilon_{n}-\epsilon_{m}},\label{eq:kubo}
\end{align}
where $\epsilon_{n}$ is the energy eigenvalue of state $|n\rangle$
, $v_{mn}^{a}(k)=\frac{1}{\hbar}\langle m|\frac{\partial H}{\partial k_{a}}|n\rangle$
are the matrix elements of the velocity operators at $a=x,y$ direction,
and $f(\epsilon)=1/\left(1+\exp(\frac{\epsilon-\mu}{k_{B}T})\right)$
is the Fermi-Dirac distribution function with $\mu$ the chemical
potential at finite temperature $T$. $k_{B}$ is the Boltzmann constant.
$\delta$ is the infinitesimal regulator. After some calculations,
the real part of the optical Hall conductivity at zero temperature
($k_{B}T=0$) and $\mu=0$ can be analytically found as (see Appendix
for details)
\begin{equation}
\mathrm{Re}\sigma_{xy}(\omega)=\frac{e^{2}}{h}\frac{1}{8\xi\tilde{\omega}}\left[2(1-4bm)\ln\left|\frac{\tilde{\omega}+\xi}{\tilde{\omega}-\xi}\right|+\sum_{s=\pm}g_{s}(\omega)\right],\label{eq:optical hall}
\end{equation}
where the dimensionless parameter $\xi=\sqrt{1-4bm+\tilde{\omega}^{2}}$
and the renormalized frequency $\tilde{\omega}=b\hbar\omega/v^{2}$,
and
\[
g_{s}(\omega)=(1-4bm-s\xi)\ln\left|\frac{\xi(1-2bm)-2b|m|\tilde{\omega}-s(1-4bm)}{\xi(1-2bm)+2b|m|\tilde{\omega}-s(1-4bm)}\right|.
\]
In the dc limit by taking $\omega\to0$, it recovers the Hall conductivity
$\sigma_{xy}(0)=\frac{1}{2}\frac{e^{2}}{h}[{\rm sgn}(m)+{\rm sgn}(b)]$
as shown in Fig. 1. In the case of $m>0$ and $b>0$, $\sigma_{xy}(0)=\pm\frac{e^{2}}{h}$
and in the case $m<0$ and $b>0$, $\sigma_{xy}(0)=0$. The blue and
yellow lines represent the two cases separately.With the frequency
increasing, the Hall conductivity deviates from the dc limit value
and becomes divergent at $\hbar\omega=mv^{2}$ due to the Rabi resonance.
This a strong indication of the existence of the band gap $mv^{2}\neq0$.
Near the region, the sign of the Hall conductivity depends on the
sign of $m$ . As the frequency further increases above the band gap
$\hbar\omega>mv^{2}$, the Hall conductivities converge to a quasi-quantized
plateau with a half-integer value $\frac{1}{2}\mathrm{sgn}(b)\frac{e^{2}}{h}$.
When the frequency is in the proper range , $mv^{2}\ll\hbar\omega\ll\frac{v^{2}}{b}$,
i.e., the dimensionless parameters $\tilde{\omega},\frac{bm}{\tilde{\omega}}\ll1$,
the real part of $\sigma_{xy}(\omega)$ is approximately,
\begin{equation}
\mathrm{Re}\sigma_{xy}(\omega)\approx\frac{e^{2}}{2h}\mathrm{sgn}(b)\left[1+\mathrm{sgn}(bm)\left(\frac{2bm}{\tilde{\omega}}\right)^{2}\right].
\end{equation}
The value of the plateau is independent of the magnitude and sign
of $m$, but is attributed by the sign of $b$, which can be regarded
as a signature of parity anomaly. $\mathrm{Re}\sigma_{xy}(\omega)$
will deviate the value of plateau if the frequency continues increasing,
\begin{equation}
\mathrm{Re}\sigma_{xy}(\omega)\approx\frac{e^{2}}{2h}\mathrm{sgn}(b)\left[1-\frac{2}{3}\tilde{\omega}^{2}\right].
\end{equation}
For comparison, we also plot $\mathrm{Re}\sigma_{xy}(\omega)$ for
the massive Dirac fermions ($m>0$ and $b=0$). The value decreases
to zero quickly after the Rabi resonance, and there is no signature
of parity anomaly.

\begin{figure}
\begin{centering}
\includegraphics[width=8cm]{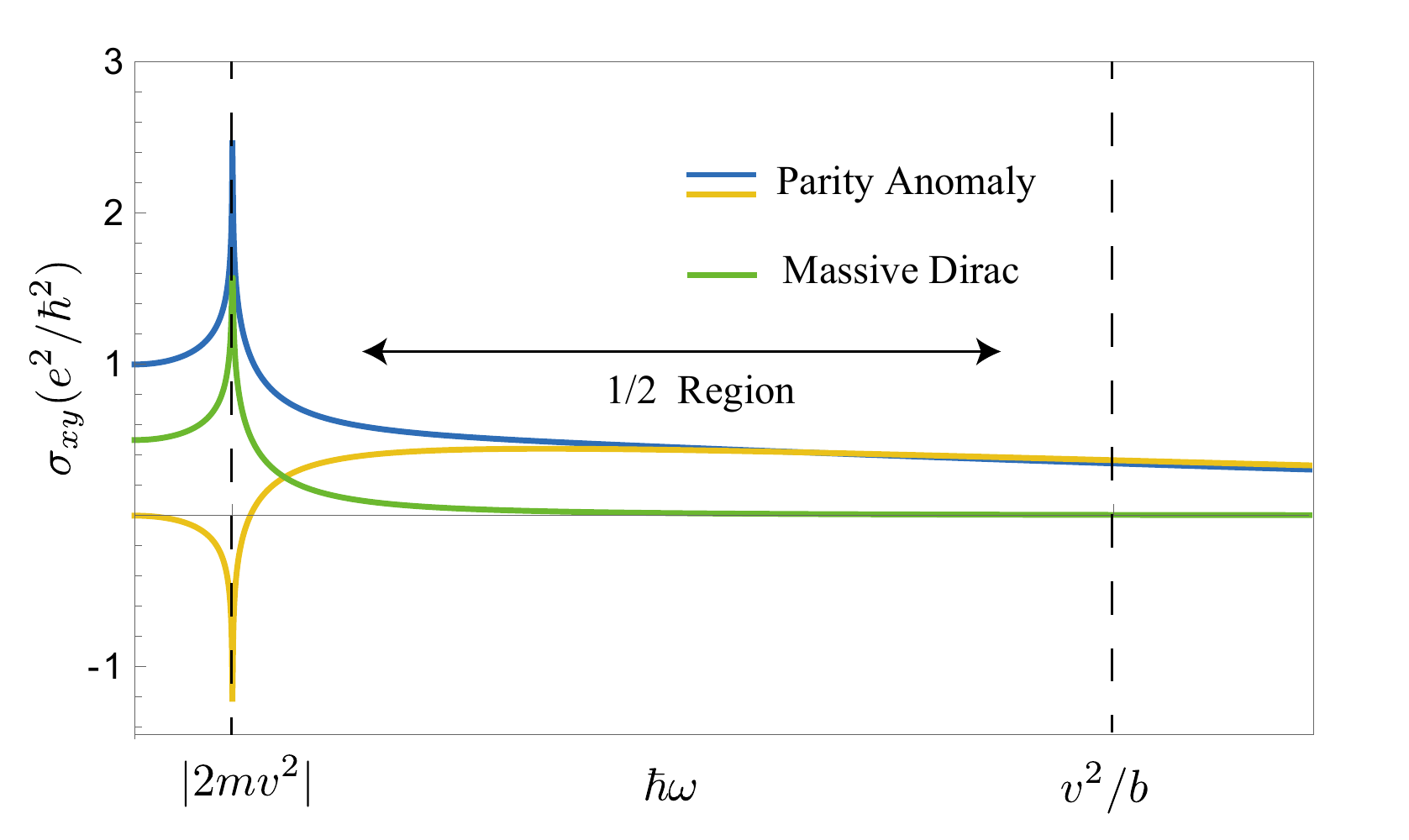}
\par\end{centering}
\caption{The comparison of optical Hall conductivity between massive Dirac
fermions and Wilson fermions. The blue and yellow lines are Wilson
fermions that are topologically non-trivial and trivial, respectively.
The green line is the massive Dirac fermion. The region sandwiched
between two dashed lines is the half-quantization region from the
parity anomaly. Here $mv^{2}=0.05\,{\rm eV}$ (green and blue line),
$-0.05\,{\rm eV}$ (yellow line), $v\hbar=0.5\,{\rm eV}\cdot\textrm{\r{A}},$
and $b\hbar^{2}=0.2\,{\rm eV}\cdot\textrm{\r{A}}^{2}$}
\end{figure}

Besides the dynamical mass, the massive Dirac fermion with mass $m$
can also be regulated by another Dirac fermion with a large Dirac
mass $M$, which is the Pauli-Villars method \citep{Dunne1999}. In
the case, the real part of the optical Hall conductivity can be found
to be 
\begin{align}
\mathrm{Re}\sigma_{xy}(\omega) & =\frac{1}{2}\frac{e^{2}}{h}\Big(\frac{mv^{2}}{\hbar\omega}\ln\left|\frac{\hbar\omega+2|m|v^{2}}{\hbar\omega-2|m|v^{2}}\right| \nonumber \\
 & +\frac{Mv^{2}}{\hbar\omega}\ln\left|\frac{\hbar\omega+2|M|v^{2}}{\hbar\omega-2|M|v^{2}}\right|\Big).
\end{align}
In the dc limit of $\omega\to0$, it is reduced to 
\begin{equation}
\mathrm{Re}\sigma_{xy}(\omega\to0)=\frac{1}{2}\frac{e^{2}}{h}[{\rm sgn}(m)+{\rm sgn}(M)].
\end{equation}
Similar to the case of Wilson fermion, the zeroth-order of Hall conductivity
is half-quantized and merely depends on the sign of the regulator
$M$. Therefore, the effect of the large mass regulator $M$ contributes
a background of half-quantized Hall conductivity and shifts the whole
curve by $\frac{1}{2}\frac{e^{2}}{h}\mathrm{sgn}(M)$.

The optical Hall conductivity will be divergent when the frequency
$\omega$ approaches the band edges $2mv^{2}$ and $2Mv^{2}$. When
$2mv^{2}\ll\hbar\omega\ll2Mv^{2}$, the real part of the optical Hall
conductivity can be expressed appropriately as following,
\begin{equation}
\sigma_{xy}\approx\frac{\mathrm{sgn}(M)}{2}\frac{e^{2}}{h}\left[1+\frac{1}{12}\left(\frac{\hbar\omega}{Mv^{2}}\right)^{2}+4{\rm sgn}(mM)\left(\frac{mv^{2}}{\hbar\omega}\right)^{2}\right].
\end{equation}
Thus in the presence of Pauli-Villars regulator, the finite-Hall conductivity
shows similar signature of parity anomaly as a half-quantized plateau.

\section{Lattice Realization}

The formulation of the lattice theory of Dirac fermion is closely
related to the Nielsen-Ninomiya no-go theorem \citep{Nielsen-81plb}.
Unlike the continuum model, the finite lattice spacing serves as a
natural UV regulator. The Wilson fermion in Eq. (\ref{eq:continuum model})
can be directly put on the lattice with no fermion doubling problem
in the presence of the regulation term $b(\hbar k)^{2}\sigma_{z}$,
which is equivalent to spin-polarized Bernevig-Hughes-Zhang (BHZ)
model \citep{Bernvig2006science}, 
\begin{align}
H_{BHZ}(k_{x},k_{y}) & =v\frac{\hbar}{a}\sum_{i=x,y}\sin(k_{i}a)\sigma_{i}\nonumber \\
 & +mv^{2}\left\{ 1-2bm\left(\frac{\hbar}{mva}\right)^{2}[2-\sum_{i=x,y}\cos(k_{i}a)]\right\} \sigma_{z},\label{eq:BHZ}
\end{align}
where $a$ is the lattice constant. The Chern number of the valence
band of this model depends on the the Dirac mass $m$ and the coefficient
$b$,
\begin{align}
C & =\frac{1}{2}\mathrm{sgn}(m)\Big\{2\mathrm{sgn}\left[1-4bm\left(\frac{\hbar}{mva}\right)^{2}\right]\nonumber \\
 & -\mathrm{sgn}\left[1-8bm\left(\frac{\hbar}{mva}\right)^{2}\right]-1\Big\}.
\end{align}
By numerically evaluating the Kubo formula in Eq. (\ref{eq:kubo}),
we obtain the optical Hall conductivity in Fig. 2 for three different
band gaps ($mv^{2}=-0.1,0.05,0.1\,\mathrm{eV}$) in Eq. (\ref{eq:BHZ})
. When $mv^{2}=0.05\mathrm{eV}$ and $mv^{2}=0.1\mathrm{eV}$, the
Chern number is $C=1$, and the optical Hall conductivities begin
with $\frac{e^{2}}{h}$ at $\omega=0$ and becomes divergent at $\hbar\omega=0.1\,\mathrm{eV}$
and $\hbar\omega=0.2\,\mathrm{eV}$, respectively. When $mv^{2}=-0.1\,{\rm eV},$
$C=0$, and the optical Hall conductivity begins with $0$ at $\omega=0$
and become divergent at $\hbar\omega=0.2\,\mathrm{eV}$. In a large
frequency regime $mv^{2}\ll\hbar\omega\ll\frac{v^{2}a^{2}}{b}$, all
the three curves approach $\frac{1}{2}\frac{e^{2}}{h}$ (as indicated
by the black line), which is consistent with the results of continuum
model in Fig. 1. In experiments, the spin-polarized BHZ models can
be realized in several two-dimensional quantum anomalous Hall effect
materials, including monolayer magnetic material $1T-\mathrm{VSe}_{2}$
\citep{Huang2021Nano}, and 2D magnetic Van der Waals heterojunction
of $\mathrm{MnNF}/\mathrm{MnNCl}$ \citep{Pan2020npj}.

\begin{figure}
\centering{}large\includegraphics[width=8cm]{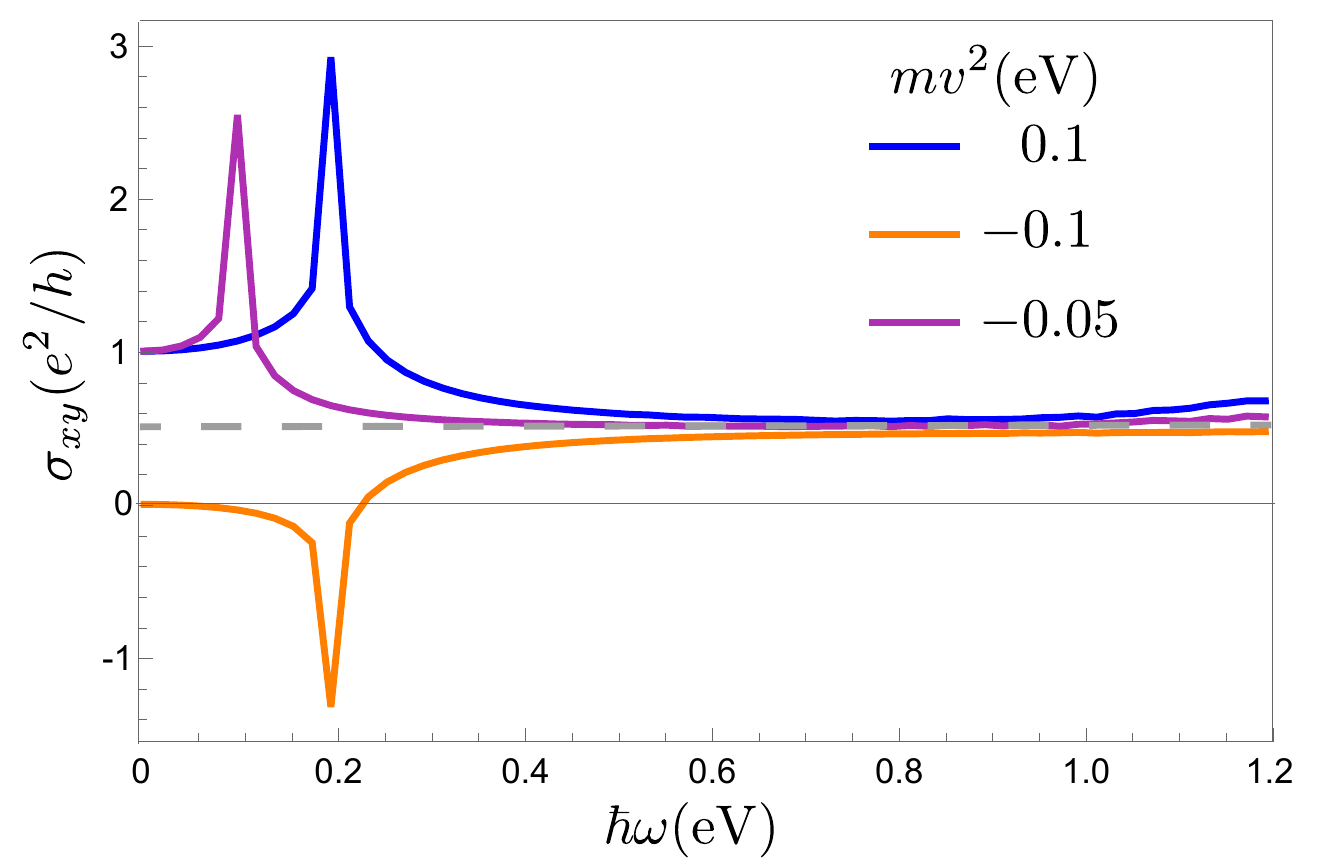}\caption{The Hall conductivity with finite-frequency of BHZ model with different
gaps. The different colors represent the different mass gap of the
model and the sign indicates the topologically trivial or non-trivial.
All samples have a plateau tending to 1/2 at the proper frequency.
The different line represent different mass show in the legends with
$v\hbar=0.5\,{\rm eV}\cdot\textrm{\r{A}},b\hbar^{2}=0.2\,{\rm eV}\cdot\textrm{\r{A}}^{2},a=1{\rm nm}.$}
\end{figure}

The first lattice model to realize parity anomaly was proposed in
the seminal paper by Haldane \citep{Haldane1988Model}. The Haldane
model can be implemented in a honeycomb lattice with the nearest hopping
$t$ and the next to nearest imaginary hopping $t'$ and an on-site
potential $M$. In terms of the Pauli matrices, the corresponding
tight-binding model can be expressed as 
\begin{equation}
H=\sum_{i=x,y,z}d_{i}\sigma_{i},
\end{equation}
where $d_{x}(\mathbf{k})=t\sum_{i}\cos(\mathbf{k}\cdot\mathbf{a}_{i})$,
$d_{y}(\mathbf{k})=t\sum_{i}\sin(\mathbf{k}\cdot\mathbf{a}_{i})$,
$d_{z}(\mathbf{k})=\Delta-2t'\sum\sin(\mathbf{k}\cdot\mathbf{b}_{i})$,
$\mathbf{a}_{i}$ is the nearest vectors of honeycomb lattice that
$\mathbf{a}_{1}=(\frac{1}{2},\frac{\sqrt{3}}{2})a$, $\mathbf{a}_{2}=(\frac{1}{2},-\frac{\sqrt{3}}{2})$a,
$\mathbf{a}_{2}=(-1,0)a$, and $\mathbf{b}_{i}=\epsilon_{ijk}(\mathbf{a}_{j}-\mathbf{a}_{k})$
with $\epsilon_{ijk}$ the anti-symmetric symbol. There are two Dirac
cones at $K$ and $K'$ valleys in the corner of the Brillouin zone,
with the band gaps $2\Delta_{+}=2\Delta+6\sqrt{3}t^{\prime}$ and
$2\Delta_{-}=2\Delta-6\sqrt{3}t^{\prime}$ respectively. When one
of the band gaps closes (e.g., $\Delta=3\sqrt{3}t'$), the low energy
theory is described by a single massless Dirac fermion with the parity
symmetry (or time-reversal symmetry). The massive Dirac fermion at
the other valley plays the role of the large mass regulator and gives
a half-quantized contribution to the Hall conductivity, $\sigma_{xy}=\frac{1}{2}\frac{e^{2}}{h}.$
Haldane thought that it is a realization of parity anomaly on the
lattice. As shown in Fig. 3(a), there are two peaks in the optical
Hall conductivity when $\hbar\omega=2\Delta_{+}$ or $\hbar\omega=2\Delta_{-}$,
and the conductivity drops to zero quickly when $\hbar\omega$ is
larger than $2\Delta_{+}$. However, when $2\Delta_{-}<\hbar\omega\ll2\Delta_{+}$,
the optical Hall conductivity is approximately half-quantized. This
condition can be realized by tuning the band gap of two valleys in
a Floquet system \citep{chen2018prb,Vogl2021}. Hence, the optical
Hall conductivity can be used to distinguish the contribution from
the low energy physics and the high energy physics (or the parity
anomaly). The Haldane model can be realized in ferromagnetic honeycomb
materials with no inversion symmetry \citep{kim2017npj,Zhou2019prl}.
Recently, some Moir\'{e} materials are reported that spontaneous magnetization
at a proper filling have a similar valley polarized quantum Hall behavior
\citep{Serlin2020sci,Stepanov2021,Sharpe19sci}. As the half-quantization
of the optical Hall conductivity does not require $\Delta_{-}=0$
exactly, it becomes more feasible in experiments.

In addition to the Haldane model, the magnetically doped topological
insulator thin film, which is the first realization of the quantum
anomalous Hall system experimentally\citep{yu2010science,Chang2013sci},
also yields two Dirac cones. Different from the Haldane model, the
two Dirac cones in topological insulators are separated in the real
space and located on the top and bottom surfaces, respectively. The
magnetic doping will open band gaps for the surface Dirac cones through
the exchange interaction \citep{liu2009prl,chen2010science}. Then,
each of the two Dirac cones contributes a one-half Hall conductivity
individually \citep{zou2022pas}. In the quantum anomalous Hall insulator
phase, the summation of two surfaces Dirac cones gives a quantized
Hall conductivity. Here we use the magnetically doped three dimensional
topological insulator model to perform the calculation, 
\begin{align}
H_{3D} & =ig(z)\alpha_{1}\alpha_{2}+v\frac{\hbar}{a}\sum_{i=x,y,z}\sin(k_{i}a)\alpha_{i}\nonumber \\
 & +m_{0}v^{2}\left\{ 1-b_{0}\left(\frac{\hbar}{m_{0}va}\right)^{2}[3-\sum_{i=x,y,z}\cos(k_{i}a)]\right\} \beta,
\end{align}
where $\alpha_{i}=\tau_{1}\sigma_{i}$ and $\beta=\tau_{3}\sigma_{0}$
are the Dirac matrices, and $a$ is the lattice constant. $v,b_{0}$
and $m_{0}$ are material parameters for a three-dimensional topological
insulator. The term $ig(z)\alpha_{1}\alpha_{2}$ is the position-dependent
Zeeman energy along the $z$-direction, which breaks the time-reversal
symmetry and generates the Dirac mass in the surface states. When
$g(z)\equiv g$ is chosen as a constant, the induced Dirac masses
of top and bottom surface states will have the same magnitude and
opposite sign. Besides, the two surface states have opposite helicity
\citep{lu2010prb}. Then, in the dc limit, each of them contributes
$\frac{e^{2}}{2h}\mathrm{sgn}(g)$ to the Hall conductivity, and the
total Hall conductivity is quantized as $e^{2}/h$. When the frequency
is nonzero, the system hosts the inversion symmetry and the optical
Hall conductivity from the two massive surface states is still identical,
as shown in Fig. 3(b). The total optical Hall conductivity of the
magnetic topological insulator thin film can be regarded as twice
of the massive Dirac fermion without any regulator. It confirms the
fact that the quantized Hall conductivity in the magnetically doped
topological insulators is determined by two gapped surface states.
This mechanism would make the behavior of optical Hall conductivity
in magnetic topological insulators different from the case of the
Haldane model. The optical Hall conductivity can be used to distinguish
multiple physical origins of quantum anomalous Hall effect in different
systems.

Moreover, if $g(z)$ is mainly localized at the top surface, it is
possible to open the gap of the top surface state only while the bottom
surface state remains gapless. It provides the most direct way to
realize the parity anomaly in experiments \citep{Mogi2022NP,zou2022pas}.
However, it is challenge to preserve a single gapless surface state.
In such a case, parity anomaly can also be detected by the optical
Hall conductivity. When there is a significant difference in the magnitude
of the band gaps for the two Dirac states, it will be similar to the
Haldane model. A half-quantized plateau can also occur if the frequency
is between the band gaps of the two Dirac cones.

\begin{figure}
\begin{centering}
\includegraphics[width=8cm]{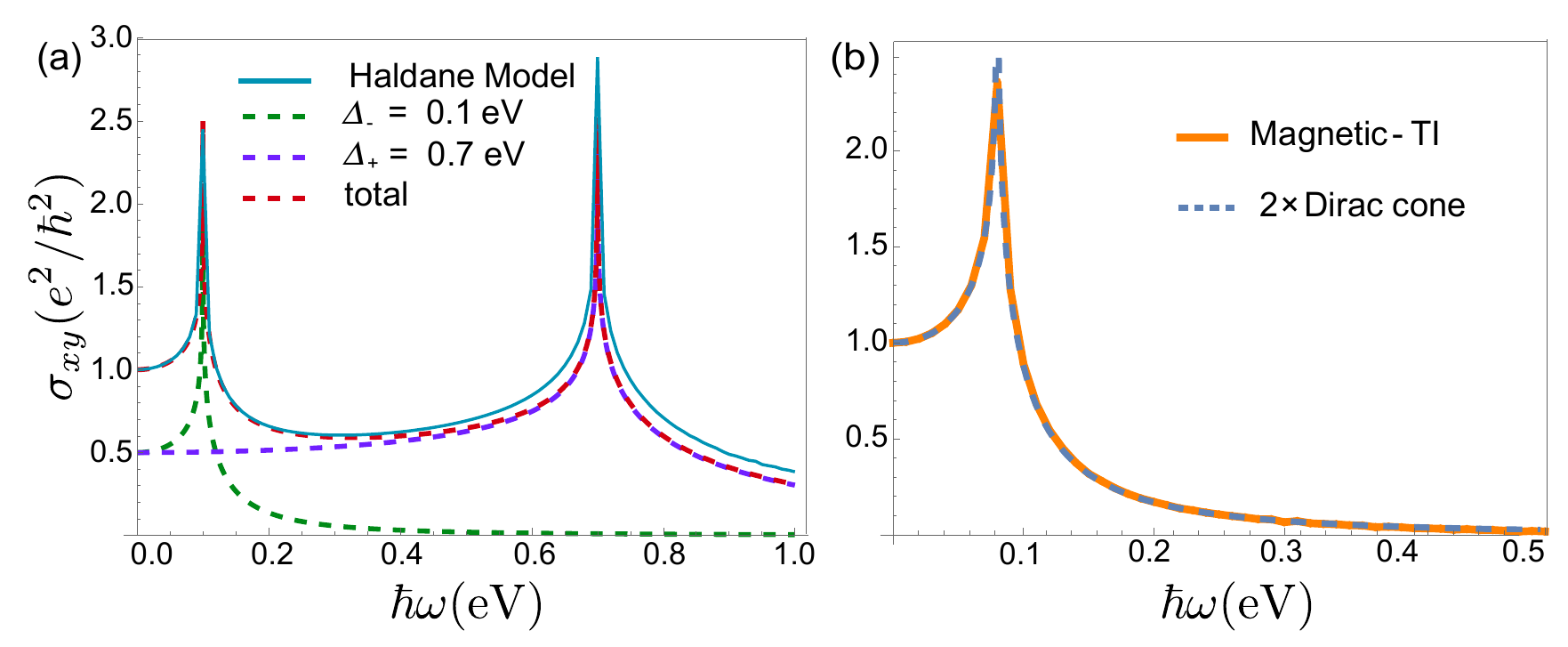}
\par\end{centering}
\caption{(a). The solid line is the optical Hall conductivity for the Haldane
model, and the dashed lines are the optical Hall conductivity for
the massive Dirac model, where the gaps equal $\Delta_{-}=0.1\,\mathrm{eV}$
and $\Delta_{+}=0.7\,\mathrm{eV}$ and hopping strength is chosen
as $t=0.5\,{\rm eV}$. (b). The optical Hall conductivity for the
magnetically doped topological insulator thin film of 8 quintuple
layer with the Zeeman term $g=0.04\,\mathrm{eV}$. The orange solid
line is the numerical result of optical Hall conductivity and the
dashed line is twice of the analytical result of a massive Dirac fermion.
The parameters for the topological insulator are chosen as $m_{0}v^{2}=0.68\,\mathrm{eV},$
$v\hbar=0.4\,\mathrm{eV}\cdot\textrm{\r{A}},$ and $b_{0}\hbar^{2}=1.4\,\mathrm{eV}\cdot\textrm{\r{A}}^{2},a=1{\rm nm}$.}
\end{figure}

\section{Experiment Implement}

In experiment, the optical Hall conductivity can be obtained by measuring
the magneto-optical effect, which reflects the information of Hall
conductivity near the sample surface.

\begin{figure}
\includegraphics[scale=0.9]{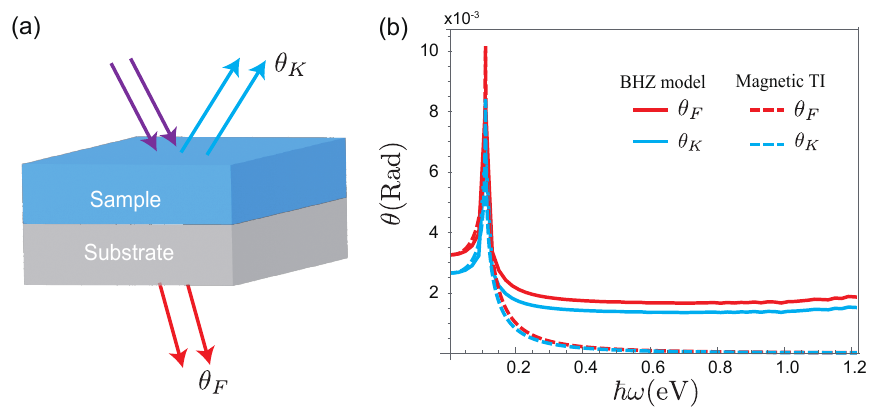}

\caption{(a).The experiment setup for the measurement of Kerr and Faraday angle.
(b). The Kerr and Faraday angle of the the BHZ model (blue and cyan
dashed lines) and the magnetic topological insulator film (red and
cyan dashed lines).The parameters of the corresponding model are the
same as the one below the Fig. 2 and Fig. 3. The refractive index
of the substrate is chosen as $n_{r}=3.46$ for InP.}
\end{figure}

As shown in figure. 4(a), we takes the sample on the substrate with
the refractive index $n_{r}>1,$ and inject a light linearly polarized
in the $x$ direction with frequency $\omega$, the electric field
of the light is ${\bf E}_{in}=E_{0}e^{i(\omega t-kx)}\hat{x}$. In
terms of electric field of transmission light ${\bf E}_{t}$ and the
reflection light as ${\bf E}_{r}$, the Kerr and Faraday angles are
defined by the $\tan\theta_{K}=E_{r}^{y}/E_{r}^{x}$ and $\tan\theta_{F}=E_{t}^{y}/E_{t}^{x}$,
respectively. The two angles can be solved by the Maxwell equation
with proper boundary condition (see Appendix B) as
\begin{align}
\tan\theta_{F}= & \frac{2\alpha\tilde{\sigma}_{xy}}{1+n_{r}+2\alpha\tilde{\sigma}_{xx}},\label{eq:faraday angle}\\
\tan\theta_{K}= & -\frac{4\alpha\tilde{\sigma}_{xy}}{n_{r}^{2}-1+4\alpha[\tilde{\sigma}_{xx}(n_{r}+\alpha\tilde{\sigma}_{xx})+\alpha\tilde{\sigma}_{xy}^{2}]},\label{eq:kerr angle}
\end{align}
where the dimensionless $\tilde{\sigma}_{xx}=\sigma_{xx}\frac{h}{e^{2}}$
and $\tilde{\sigma}_{xy}=\sigma_{xy}\frac{h}{e^{2}}$ are the dimensionless
transverse conductivity and the Hall conductivity, respectively, and
$\alpha\approx1/137$ is the fine structure constant. When $\tilde{\sigma}_{xx}$
and $\tilde{\sigma}_{xy}$ are much smaller than $\alpha^{-1}$ and
the refractive index of the substrate $n_{r}>1$, the two angles can
be approximated as $\theta_{F}=\arctan\frac{2\alpha}{1+n_{r}}\tilde{\sigma}_{xy}$
and $\theta_{K}=-\arctan\frac{4\alpha}{n_{r}^{2}-1}\tilde{\sigma}_{xy}$,
respectively. In Fig. 4(b), we plot $\theta_{F}$ and $-\theta_{K}$
as functions of $\omega$ for the magnetic topological insulator and
BHZ model. We consider the case that the wavelength of light is much
larger than the thickness of the sample i.e., $d\ll\lambda$ and the
insulating subtract is InP with $n_{r}=3.46$ \citep{Mogi2022NP}.
In this limit, the bottom and top surfaces can be viewed as a whole.
Therefore, $\text{\ensuremath{\sigma_{xx}} }$ and $\sigma_{xy}$
in Eq. (\ref{eq:faraday angle}) and (\ref{eq:kerr angle}) denote
the total longitudinal and Hall conductivities from two surfaces,
respectively. At zero frequency, the chemical potential inside Dirac
gap of both surfaces, we have $\sigma_{xx}=0$ and $\sigma_{xy}=e^{2}/h$,
the Faraday and Kerr angles have universal values $\theta_{F}\approx3.28\times10^{-3}\,{\rm rad}$
and $\theta_{K}\approx-2.66\times10^{-3}\,{\rm rad}$.. At finite
frequencies, the $\theta_{F}$ and $\theta_{K}$ for the BHZ model
display a plateau with half of the zero frequency value for $\hbar\omega\gg2|m|v^{2}$.
As indicated by the red and cyan dashed lines in Fig. 4(b), the Kerr
and Faraday angles of magnetic topological insulator film drop to
zero after $\hbar\omega\gg2g$, which display the same character as
the Hall conductivity in Fig. 3(b).

Furthermore, the optical Hall conductivity can be deduced from the
absorption rate of the circularly polarized light $\Gamma_{\pm}(\omega)$
\citep{Tran2017SA}. The imaginary part of optical Hall conductivity
can be related to the difference between the absorption rate of the
left-hand and the right-hand light as $\mathrm{Im}\sigma_{xy}(\omega)=\hbar\omega(\Gamma_{+}-\Gamma_{-})/(8AE^{2}),$
where $E$ is the intensity of light and $A$ is the area of the sample.
From the Kramers-Kronig relation, the real part of the optical Hall
conductivity can be obtained as $\mathrm{Re}\sigma_{xy}(\omega)=\frac{2}{\pi}{\cal P}\int_{0}^{\infty}\frac{\omega'\mathrm{Im}\sigma_{xy}(\omega')d\omega'}{\omega'^{2}-\omega^{2}}$
with $\mathcal{P}$ denoting the Cauchy principal value, and the half-quantized
plateau can be found at finite-frequency in the $\mathrm{Re}\sigma_{xy}.$

\section{Discussion and summary}

Besides the light frequency, disorder and temperature can also be
used to smear off the low energy contribution in the Hall conductivity
and leave the parity anomaly contribution from the high energy only
\citep{Tutschku2020prb}. Therefore, we can expect a similar half-quantized
plateau in the Hall conductivity at finite temperature or finite disorder.
For the disordered system, using the Born approximation, the disorder
effect can be introduced phenomenologically by the quasi-particle
self-energy $\Gamma$ in the Green function, i.e. $G^{r/a}=(i\omega_{n}-H\pm i\Gamma)^{-1}.$
Thus, the Hall conductivity in the presence of disorder can be expressed
as
\begin{equation}
\sigma_{xy}=e^{2}\hbar\int\frac{d^{2}k}{(2\pi)^{2}}\sum\frac{\mathrm{Im}(v_{mn}^{x}v_{nm}^{y})[f(\epsilon_{m})-f(\epsilon_{n})]}{(\epsilon_{m}-\epsilon_{n})^{2}+\Gamma^{2}}.
\end{equation}
Compared to the Kubo formula at finite-frequency, $\sigma_{xy}$ can
be obtained by taking an Analytic continuation that replace the frequency
$\hbar\omega$ with $i\Gamma$. When $mv^{2}\ll\Gamma\ll\frac{v^{2}}{b},$
The asymptotic behavior of Hall conductivity reads 
\begin{equation}
\mathrm{Re}\sigma_{xy}(\Gamma)=\frac{e^{2}}{2h}{\rm sgn}(b)\left[1+\frac{2}{3}\left(\frac{b\Gamma}{v^{2}}\right)^{2}-\mathrm{sgn}(mb)\left(\frac{2mv^{2}}{\Gamma}\right)^{2}\right],
\end{equation}
 where the leading order is a half-quantized value $\frac{1}{2}{\rm sgn}(b)$.

In addition to disorder in the system, the temperature can also lead
to the half-quantized plateau when the temperature $k_{B}T$ satisfies
that $mv^{2}\ll k_{B}T\ll\frac{v^{2}}{b}.$ The effect of the temperature
can be seen as averaging the Berry curvature of the conduction band
and valence band near the Fermi surface, which can erase the contribution
of the low energy part, and only the contribution from the high energy
part remains. This also separates the parity anomaly part and allows
the system to show the half-quantized plateau. We plot the Hall conductivity
as a function of temperature $k_{B}T$ and the self-energy $\Gamma$
in the same dimensionless energy scale in Fig.5. The Hall conductivity
shows a similar nearly half-quantized feature as the optical Hall
conductivity in Fig.1.

\begin{figure}
\includegraphics[scale=0.6]{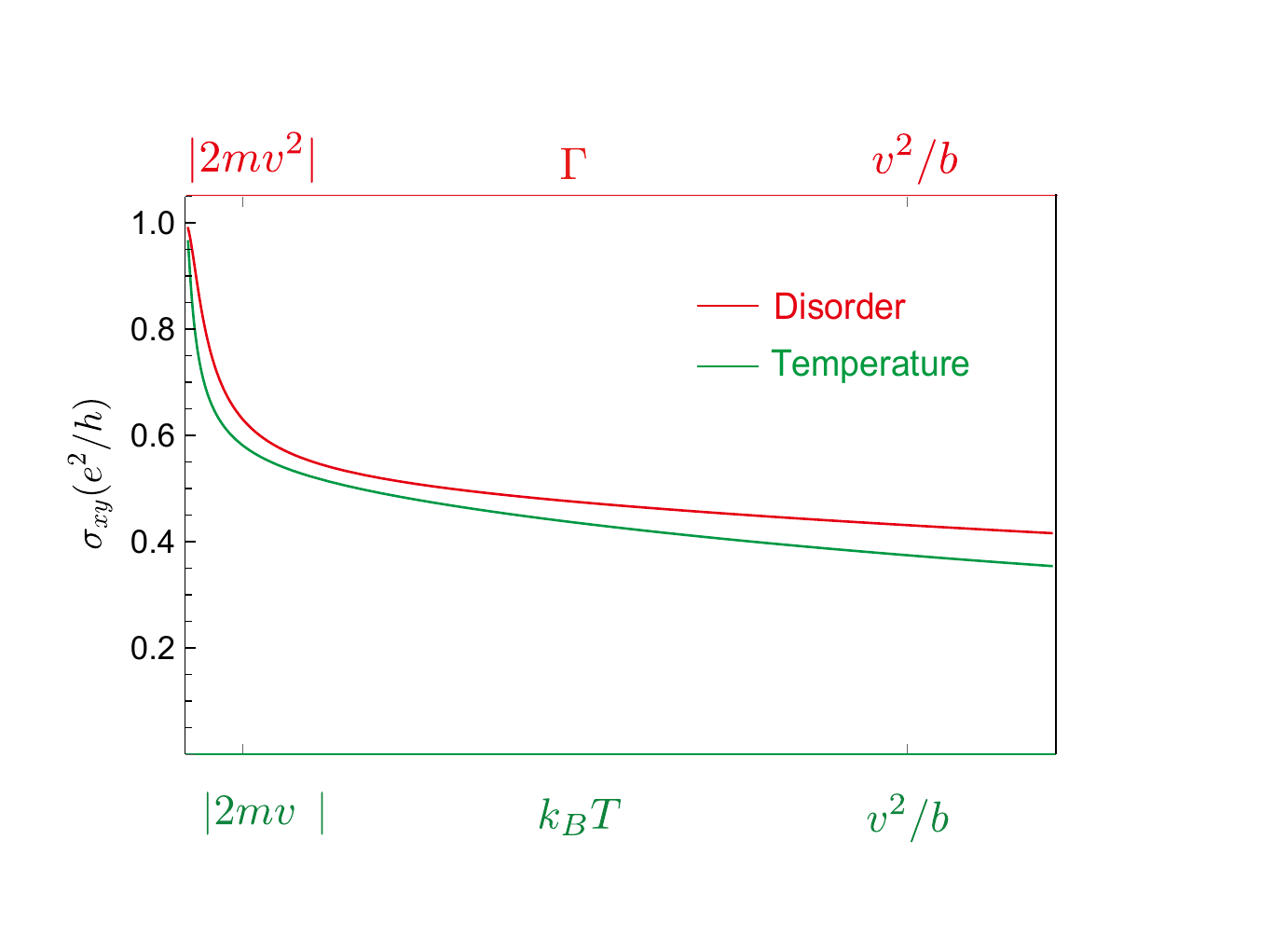}

\caption{The Hall conductivity of Wilson fermion as functions of temperature(pink
line) and disorder(cyan line), the temperature $k_{B}T$ and the disorder
$\Gamma$ have been renormalized into the same dimensionless energy
scale. The parameters are $v\hbar=0.5\,\mathrm{eV}\cdot\textrm{\r{A}},b\hbar^{2}=0.2\,\mathrm{eV}\cdot\textrm{\r{A}}^{2},mv^{2}=0.05\,{\rm eV}.$}
\end{figure}

In quantum anomalous Hall systems, the quantized Hall conductivity
usually consists of two parts, one is the contribution from massive
Dirac cone and the other is the parity anomaly contribution in the
high energy region. In this article, the finite-frequency method is
introduced to separate these two different sources of Hall conductivity.
When the applied frequency is in the proper interval, the Hall conductivity
of the system exhibits a half-quantized plateau, and this plateau
is considered a direct manifestation of the parity anomaly. Thus in
the condensed matter system, the half-quantized optical Hall conductivity
can be used as a marker to characterize the parity anomaly.

\section*{Acknowledgments}

This work was supported by the National Key R\&D Program of China
under Grant No. 2019YFA0308603 and the Research Grants Council, University
Grants Committee, Hong Kong under Grant No. C7012-21G and No. 17301220.

\appendix

\section{Calculation of Optical Hall conductivity}

For a general $2\times2$ matrix Hamiltonian in the form $H={\bf d}\cdot\sigma$
with $\mathbf{d}=(d_{x},d_{y},d_{z})$, the eigenenergy and eigenstates
can be found as 
\begin{align*}
E_{\chi} & =\chi\sqrt{d_{x}^{2}+d_{y}^{2}+d_{z}^{2}}=\chi d,\\
|\phi_{\chi}\rangle & =\begin{pmatrix}\frac{\hat{d}_{x}-i\hat{d}_{y}}{\sqrt{2(1-\chi\hat{d}_{z})}}\\
\frac{1-\hat{d}_{z}}{\sqrt{2(1-\chi\hat{d}_{z})}}
\end{pmatrix},
\end{align*}
where $\chi=+$ is for the conduction band and $\chi=-1$ is for the
valence band. $\hat{d_{i}}$ with $i=x,y,z$ are defined as $\hat{d_{i}}=d_{i}/d$.

In the eigen energy basis, the matrix elements of the velocity operators
are defined as 
\begin{align*}
v_{\chi\chi^{\prime}}^{x} & =\frac{1}{\hbar}\sum_{i=x,y,z}\frac{\partial d_{i}}{\partial k_{x}}\langle\phi_{\chi}|\sigma_{i}|\phi_{\chi^{\prime}}\rangle,\\
v_{\chi\chi^{\prime}}^{y} & =\frac{1}{\hbar}\sum_{i=x,y,z}\frac{\partial d_{i}}{\partial k_{y}}\langle\phi_{\chi}|\sigma_{i}|\phi_{\chi^{\prime}}\rangle.
\end{align*}
After a straightforward calculation, one can obtain the imaginary
part of the velocity product as 
\[
{\rm Im}(v_{\chi\bar{\chi}}^{x}v_{\bar{\chi}\chi}^{y})=\chi\frac{d^{2}}{\hbar^{2}}\hat{\mathbf{d}}\cdot(\partial_{x}\hat{\mathbf{d}}\times\partial_{y}\hat{\mathbf{d}}).
\]
 For the BHZ model, $\mathbf{d}=(v\hbar k_{x},v\hbar k_{y},mv^{2}-bk^{2}\hbar^{2})$
and $d=\sqrt{v^{2}\hbar^{2}k^{2}+(mv^{2}-bk^{2}\hbar^{2})^{2}}$,
we have ${\rm Im}v_{\chi\bar{\chi}}^{x}v_{\bar{\chi}\chi}^{y}=\chi\frac{v^{2}}{d}(mv^{2}+b\hbar^{2}k^{2})$.
As a result, the real part of the optical Hall conductivity at zero
temperature in Eq. (\ref{eq:kubo}) is given by
\begin{align}
&{\rm Re}\sigma_{xy}(\omega)  =\frac{e^{2}}{h}\frac{1}{8\xi\tilde{\omega}}\sum_{s}\left(1-4bm+s\xi\right)\nonumber \\
 & \times\mathrm{arcoth}\left.\left(\frac{2\frac{b\hbar k}{v}\tilde{\omega}\sqrt{1+(\frac{mv}{\hbar k}-\frac{b\hbar k}{v})^{2}}}{s(1-4bm)+\xi\left(1-2bm+2\left(\frac{b\hbar k}{v}\right)^{2}\right)}\right)\right|_{0}^{+\infty}.
\end{align}
Here $\xi$ and $\tilde{\omega}$ is as the same definition as in
the main text. Using the analytical continuation $\mathrm{arcoth}(x)=\frac{1}{2}\left[\ln\frac{x+1}{x}-\ln\frac{x-1}{x}\right]$,
we obtain Eq. (\ref{eq:optical hall}) in Sec. III.

\section{Kerr and Faraday angle}

For the structure in the Sec. V, Faraday's law of induction tells
\begin{align*}
\nabla\times\mathbf{E} & =-\mu_{0}\frac{\partial\mathbf{H}}{\partial t}.
\end{align*}
Here we assume that the thickness of the sample $d$ is much smaller
than the wave-length of the light such that the sample can be regarded
as a two-dimensional system and the boundary condition can be written
as following

\begin{align*}
\mathbf{E} & =\mathbf{E}_{0}-\mathbf{E}_{r}+\mathbf{E}_{t};\\
\mathbf{H} & =\mathbf{H}_{0}-\mathbf{H}_{r}+\mathbf{H}_{t};\\
\hat{\mathbf{z}}\times\mathbf{E} & =0;\\
\hat{\mathbf{z}}\times\mathbf{H} & =4\pi\mathbf{j}.
\end{align*}
Here $\mathbf{E}_{0}=E_{0}e^{i\omega t-kz}\hat{\mathbf{x}}$ is the
incidentl light, $\mathbf{E}_{r},\mathbf{H}_{r}$ is the reflection
light and $\mathbf{E}_{t},\mathbf{H}_{t}$ is the transmission light.
Using the Maxwell equations, we have $H_{i}=-\sqrt{\frac{\epsilon_{i}}{\mu_{i}}}i\tau_{y}E_{i}$
for $i=0,t$ and $H_{r}=\sqrt{\frac{\epsilon_{r}}{\mu_{r}}}i\tau_{y}E_{r}.$
The current only depends on the transmission electric field as$j=\sigma(\omega)E_{t}$.

The boundary condition for the electric field can be written as 
\begin{align*}
-E_{r,x}+E_{t,x}-E_{0,x} & =0;\\
-\sqrt{\frac{\epsilon_{0}}{\mu_{0}}}E_{r,y}-\sqrt{\frac{\epsilon_{t}}{\mu_{t}}}E_{t,y}+\sqrt{\frac{\epsilon_{0}}{\mu_{0}}}E_{0,y} & =4\pi\left(\sigma_{yy}E_{t,y}-E_{t,x}\sigma_{xy}\right);\\
-E_{r,y}+E_{t,y}-E_{0,y} & =0;\\
\sqrt{\frac{\epsilon_{0}}{\mu_{0}}}E_{r,x}+\sqrt{\frac{\epsilon_{t}}{\mu_{t}}}E_{t,x}-\sqrt{\frac{\epsilon_{t}}{\mu_{t}}}E_{0,x} & =-4\pi\left(\sigma_{xx}E_{t,x}+E_{t,y}\sigma_{xy}\right).
\end{align*}
Set $E_{0,y}=0$ and $E_{0,x}=E_{0}$. We suppose that the system
has the symmetry that $\sigma_{xx}=\sigma_{yy}$ and $\sigma_{yx}=-\sigma_{xy}$.
In this way, we can find that

\begin{align*}
E_{t,x} & =\frac{2E_{0}\sqrt{\frac{\epsilon_{0}}{\mu_{0}}}\left(4\pi\sigma_{xx}+\sqrt{\frac{\epsilon_{t}}{\mu_{t}}}+\sqrt{\frac{\epsilon_{0}}{\mu_{0}}}\right)}{\left(4\pi\sigma_{xx}+\sqrt{\frac{\epsilon_{t}}{\mu_{t}}}+\sqrt{\frac{\epsilon_{0}}{\mu_{0}}}\right)^{2}+16\pi^{2}\sigma_{xy}^{2}};\\
E_{t,y} & =\frac{8\pi E_{0}\sqrt{\frac{\epsilon_{0}}{\mu_{0}}}\sigma_{xy}}{\left(4\pi\sigma_{xx}+\sqrt{\frac{\epsilon_{t}}{\mu_{t}}}+\sqrt{\frac{\epsilon_{0}}{\mu_{0}}}\right)^{2}+16\pi^{2}\sigma_{xy}^{2}};\\
E_{r,x} & =\frac{-E_{0}\left((4\pi\sigma_{xx}+\sqrt{\frac{\epsilon_{t}}{\mu_{t}}})^{2}+16\pi^{2}\sigma_{xy}^{2}-\frac{\epsilon_{0}}{\mu_{0}}\right)}{\left(4\pi\sigma_{xx}+\sqrt{\frac{\epsilon_{t}}{\mu_{t}}}+\sqrt{\frac{\epsilon_{0}}{\mu_{0}}}\right)^{2}+16\pi^{2}\sigma_{xy}^{2}};\\
E_{r,y} & =\frac{8\pi E_{0}\sqrt{\frac{\epsilon_{0}}{\mu_{0}}}\sigma_{xy}}{\left(4\pi\sigma_{xx}+\sqrt{\frac{\epsilon_{t}}{\mu_{t}}}+\sqrt{\frac{\epsilon_{0}}{\mu_{0}}}\right)^{2}+16\pi^{2}\sigma_{xy}^{2}}.
\end{align*}
The Faraday angle $\theta_{F}$ and Kerr angle $\theta_{K}$ are 
\[
\tan\theta_{F}=\frac{E_{t,y}}{E_{t,x}}=\frac{4\pi\sigma_{xy}}{\sqrt{\frac{\epsilon_{0}}{\mu_{0}}}+\sqrt{\frac{\epsilon_{t}}{\mu_{t}}}+4\pi\sigma_{xx}}
\]
and 
\[
\tan\theta_{K}=-\frac{E_{r,y}}{E_{r,x}}=-\frac{8\pi\sqrt{\frac{\epsilon_{0}}{\mu_{0}}}\sigma_{xy}}{(4\pi\sigma_{xx}+\sqrt{\frac{\epsilon_{t}}{\mu_{t}}})^{2}+16\pi^{2}\sigma_{xy}^{2}-\frac{\epsilon_{0}}{\mu_{0}}}.
\]
Using the fine structure constant $\alpha=\frac{2\pi e^{2}}{h}\sqrt{\frac{\mu_{0}}{\epsilon_{0}}}$,
we obtain the Eqs. (\ref{eq:faraday angle}) and (\ref{eq:kerr angle})
in the Sec. V.

\end{document}